\documentclass[a4paper,12pt,english]{article}

\usepackage{amsfonts,bm,amssymb,euscript,array,babel,cite}
\usepackage{mathtools}
\usepackage{tikz-cd}
\usepackage{amsmath}
\usepackage{tensor}
\usepackage{hhline}
\usepackage[amsmath]{empheq}
\usepackage{hyperref,multirow}
\usepackage{cancel}
\usepackage{mathrsfs}
\usepackage{pdfsync}
\usepackage{tikz}

\usepackage{framed}

\newsavebox{\mysaveboxM}
\newsavebox{\mysaveboxT}

\makeatletter

\newcommand{\dd}{\mathrm{d}}
\newcommand{\DD}{\mathrm{D}}

\newcommand{\w}{\wedge}

\newcommand{\be}{\begin{equation}}
\newcommand{\ee}{\end{equation}}

\def\nn{\nonumber}
\def \bea{\begin{eqnarray}} 
\def\eea{\end{eqnarray}}
\newcommand{\mf}{\mathfrak}
\def\mc{\mathcal}

\def\bi{\begin{itemize}} 
\def\ei{\end{itemize}}

\makeatother

\def\a{\alpha} \def\b{\beta} \def\g{\gamma} \def\G{\Gamma} \def\d{\delta} 
\def\e{\epsilon} 
   
  \def\m{\mu}
\def\n{\nu}    \def\r{\rho}
\def\s{\sigma} \def\S{\Sigma}

\def\one{\mbox{1 \kern-.59em {\rm l}}}

\numberwithin{equation}{section}

\begin{document}

\makeatother
\parindent=0cm
\renewcommand{\title}[1]{\vspace{10mm}\noindent{\Large{\bf #1}}\vspace{8mm}} \newcommand{\authors}[1]{\noindent{\large #1}\vspace{5mm}} \newcommand{\address}[1]{{\itshape #1\vspace{2mm}}}

\begin{titlepage}

\begin{flushright}
	RBI-ThPhys-2022-25
\end{flushright}

\begin{center}

\title{ {\large {Topological Dirac Sigma Models \\[5pt] \& the Classical Master Equation}}}

  \authors{\small{Athanasios {Chatzistavrakidis}{$^{\a}$}, Larisa Jonke{$^{\a,\beta}$}, Thomas Strobl$^{\gamma}$ \& Grgur \v{S}imuni\'c{$^{\a}$}}
    }
 
% \vskip 2mm
 
  \address{ $^{\a}$ Division of Theoretical Physics, Rudjer Bo\v skovi\'c Institute \\ Bijeni\v cka 54, 10000 Zagreb, Croatia \\[5pt]
 
 $^{\beta}$ School of Theoretical Physics,  Dublin Institute for Advanced Studies \\
 10 Burlington Road,  Dublin 4,  Ireland \\[5pt]
 
 $^{\gamma}$ Institut Camille Jordan, Universit\'e Claude Bernard Lyon 1 \\ 
 43 boulevard du 11 novembre 1918, 69622 Villeurbanne cedex, France

 }

\vskip 2cm

\begin{abstract}

We present the construction of the classical Batalin-Vilkovisky action for topological Dirac sigma models. The latter are two-dimensional topological field theories that simultaneously generalise the completely gauged Wess-Zumino-Novikov-Witten model and the Poisson sigma model. Their underlying structure is that of Dirac manifolds associated to maximal isotropic and integrable subbundles of an exact Courant algebroid twisted by a 3-form. In contrast to the Poisson sigma model, the AKSZ construction is not applicable for the general Dirac sigma model. We therefore follow a direct approach for determining a suitable BV extension of the classical action functional with ghosts and antifields satisfying the classical master equation. Special attention is paid on target space covariance, which requires the introduction of two connections with torsion on the Dirac structure.   

\end{abstract}

\end{center}

\vskip 2cm

\end{titlepage}

\setcounter{footnote}{0}
\tableofcontents

%\newpage

\section{Introduction}
\label{sec1}

Two-dimensional sigma models play a central role in various physical contexts, such as conformal field theory, string theory and certain problems in condensed matter physics. A particular class of such theories consists of those whose moduli space of classical solutions modulo gauge transformations is finite-dimensional, in which case they are referred as topological or almost topological field theories.\footnote{A theory is called topological if, up to gauge transformations, the classical solutions do not depend on background data of the source manifold, but just on its topology---or likewise, on the quantum level, if expectation values of gauge invariant quantities have such a feature.} Well known examples of topological field theories in two dimensions include the A/B topological string models \cite{Witten:1988xj,Witten:1991zz}, the $G/G$ Wess-Zumino-Novikov-Witten  (WZNW) model \cite{Alekseev:1995py} and the Poisson sigma model \cite{Ikeda,SchallerStrobl}, the two-dimensional Yang-Mills theory is an example of an almost topological theory. 

Notably, the Poisson sigma model is obtained as the first instance in those class of theories whose Batalin-Vilkovisky (BV) action can be found using the celebrated AKSZ construction \cite{Alexandrov:1995kv}. This is a powerful geometric approach to the BV quantization of gauge theories and, more generally, constrained Hamiltonian systems. Its geometric backbone is a differential graded manifold equipped with a compatible graded symplectic structure of degree $n$, also known as QP$n$ manifold, which plays the role of the target space of the BV formulation of the sigma model. The case of $n=1$ corresponds to the Poisson sigma model through the associated Lie algebroid structure on the cotangent bundle of the underlying manifold. The authors of Ref. \cite{Alexandrov:1995kv} then show that its gauge fixed version is precisely the topological A model. 
This gives a precise relation between the Poisson sigma model and the A model.

The aforementioned theories certainly do not exhaust the possibilities for two-dimensional topological field theories. Another interesting class was found by exploring the relation between topological WZNW models and the Poisson sigma model \cite{Kotov:2004wz}. This led to the construction of the so-called Dirac sigma models. These are two-dimensional topological field theories whose underlying geometrical structure is that of Dirac manifolds \cite{dirac} in much the same way as Poisson manifolds underlie Poisson sigma models. From another viewpoint, more akin to gauge theory, topological Dirac sigma models are associated to a certain class of Lie algebroids, Dirac structures, which are obtained as maximal isotropic and involutive subbundles of an exact Courant algebroid \cite{Salnikov:2013pwa,ChatzistavrakidisAHP}. By construction, Dirac sigma models involve a Wess-Zumino term using a closed 3-form $H$. Since the cotangent bundle may be given the structure of such a Lie algebroid, a special case of Dirac sigma models is the $H$-twisted 
Poisson sigma model, first considered in \cite{Klimcik:2001vg}, with underlying structure that of a twisted Poisson manifold \cite{SeveraWeinstein}.  

The presence of the Wess-Zumino term in Dirac sigma models in general and in the $H$-twisted Poisson sigma model in particular has an interesting consequence. Although one can always determine a Q-structure (a (co)homological vector field) on the target space manifold, a (graded symplectic) P-structure is not available in general and when it is, in the case of the twisted Poisson sigma model, it is not compatible with the Q-structure and as a consequence the target space does not have the structure of a QP manifold. In the latter case, the obstruction is solely due to the 3-form $H$ that gives rise to the Wess-Zumino term. This in turn means that the AKSZ construction cannot be applied directly to determine the BV action of the model. This is one of the reasons that although the BV action of the Poisson sigma model was essentially found already in \cite{Alexandrov:1995kv}, the one of its $H$-twisted version was found only recently in \cite{Ikeda:2019czt}. On the other hand, for the more general class of topological Dirac sigma models, which include the twisted Poisson sigma model as a special case, the BV action remains unknown. 
The main purpose of this paper is to find this action.

In order to determine the BV action for Dirac sigma models, we follow the traditional BV/BRST method with an additional geometric touch in the philosophy of ``covariantization of symmetries''. Specifically, after introducing the ghost fields we define the BRST operator and test its action on all fields and ghosts. (The theories we consider are irreducible in the sense of e.g. \cite{HT} and therefore there are no ghosts for ghosts, at least when the worldsheet of the theory has trivial topology and the boundary conditions are chosen appropriately). We find that it is only nilpotent on-shell, in other words, it does not square to zero without taking into account the classical equations of motion for the model. This necessitates the introduction of antifields and antighosts. As usual, there exists a BV ``antibracket'' $(\cdot,\cdot)_{\text{BV}}$ in the space of fields/ghosts and antifields/antighosts and the extension of the classical action functional to a BV functional should be such that the extended one, say ${\cal S}_{\text{BV}}$, satisfies the classical master equation
\be 
({\cal S}_{\text{BV}},{\cal S}_{\text{BV}})_{\text{BV}}=0\,.
\ee 
We show that ${\cal S}_{\text{BV}}$ has three contributions, with 0, 1 and 2 antifields, respectively. 

An important piece in our analysis is the introduction of two vector bundle connections. The role of these connections is to covariantize the symmetries of the model so that the gauge transformations, the classical field equations and the action functional itself contain tensorial quantities and they are defined not only on a local patch but in an intrinsic, basis-independent way. These connections were determined in a closed form for the topological version of Dirac sigma models already in Ref. \cite{Chatzistavrakidis:2016jci}. In the special case of the $H$-twisted Poisson sigma model they turn out to be identical, and in effect there is only a single connection then, which is not the case in general. The importance of this ingredient in the construction of the BV action lies in the fact that the terms in ${\cal S}_{\text{BV}}$ quadratic in the antifields are proportional to the basic curvatures associated to each connection, measuring how much each connection deviates from being compatible with the Lie algebroid structure on the corresponding vector bundle. This is a direct generalization of the results of \cite{Ikeda:2019czt} for the $H$-twisted Poisson sigma model and its BV action with one connection. 

There is still one important remark to make: There is also a non-topological version of the Dirac sigma model \cite{Kotov:2014dqa}, where a Dirac structure is replaced by only an involutive and isotropic subset of the sections in an exact Courant algebroid---in the case it is given by a maximally isotropic subbundle one obtains the topological ones of this paper as a special case; in the other extreme case, where it is just zero, one obtains a non-gauged standard sigma model with Wess-Zumino term. The BV formulation of the non-topological Dirac sigma models remains an open problem as of now. 

The rest of the paper is organised as follows. 
In Section \ref{sec2} we first give a brief overview of (also non-topological) Dirac sigma models in the approach of nonlinear gauge theory and their relation to generalised geometry, emphasizing the need and role of the two vector bundle connections. We highlight the different torsion and curvature tensors that appear in the problem, paying attention in distinguishing the ones on the target space manifold and the ones on the vector bundle defined over it. In the spirit of the covariantization philosophy, we discuss the role of the two connections in finding the target space covariant form of the field equations and gauge transformations. In Section \ref{sec3},  we first present the main argument for inapplicability of the AKSZ construction for the models under study. Then we determine the BV action for Dirac sigma models using the traditional BV formalism and express it in a covariant form. Specifically, first we compute the square of the BRST operator on fields and ghosts, which leads us to the introduction of antifields and the antibracket in the full field space. We then extend the classical action with suitable terms linear and quadratic in the antifields and show that the classical master equation is satisfied. Section \ref{sec4} contains our conclusions and outlook to future work and the brief appendix \ref{appa} summarizes our conventions.

\section{Dirac Sigma Models and Generalised Geometry}  
\label{sec2}

\subsection{Dirac Sigma Models as 2D gauge theory}
\label{sec21}

One way to introduce  Dirac sigma models is through nonlinear gauge theory in two dimensions. The starting point is a scalar field theory where the fields $X^{\mu}, \mu=1,\dots, d$ are components of a sigma model map $X\colon \S\to M$ from the 2D spacetime $\S$, the worldsheet, to a $d$-dimensional target space $M$. The corresponding action functional has the general form 
\be \label{sigmamodel}
S[X]=-\int_{\S} \left(\frac 12\, g_{\m\n}(X)\dd X^{\mu}\w\ast\, \dd X^{\n}+\frac 12 B_{\m\n}(X)\dd X^{\mu}\w\dd X^{\nu}\right)-\int_{\widehat{\Sigma}} X^{\ast}H\,.
\ee    
 When we refer to local coordinates in $\S$ and $M$, they will be denoted as $\s^{\a}, \a=0,1$ and $x^{\m}, \m=1,\dots,d$ respectively, in which case $X^{\m}=X^{\ast}(x^{\mu})$, where $X^{\ast}$ is the pull-back map corresponding to $X$. In this paper, we consider $\S$ to be equipped with  the 2D Minkowski metric $\eta=(\eta_{\a\b})$ with signature $(-1,1)$; it enters the functional \eqref{sigmamodel} only by means of the Hodge duality operator $\ast$. 
 The component form of this action and further conventions in two dimensions are collected in Appendix \ref{appa}. 

The field-dependent couplings $g_{\m\n}$ and $B_{\m\n}$ of the theory are   
pullbacks of a Riemannian metric $g$ and a 2-form $B$ on the target space, respectively, which we denote by the same symbols to avoid clutter; namely $g_{\m\n}(X)=X^{\ast}g_{\m\n}(x)$ and similarly for $B_{\m\n}$. Finally, $H$ is a closed 3-form on $M$ whose pullback introduces a Wess-Zumino coupling to the 2D field theory, supported on an open membrane $\widehat{\Sigma}$ whose boundary is the worldsheet $\S$. As usual, $H$ is not necessarily an exact 3-form and the path integral of the theory is meaningful as long as it defines an integral cohomology class, in which case the exponential of the Wess-Zumino term does not depend on the choice of $\widehat{\Sigma}$ \cite{Witten:1983ar}. In the following, we use this fact to absorb the 2D topological term of \eqref{sigmamodel} in the Wess-Zumino term and therefore $B_{\mu\nu}$ will not appear in our analysis explicitly, but only through exact contributions stemming from $H$. Thus, essentially the background fields of the theory are $(g,H)$, in other words a generalised metric, see e.g. Ref. \cite{Severa:2019ddq}.

Having at hand the action functional \eqref{sigmamodel}, one may ask under which conditions for the geometrical data $(g,H)$ there exists an extension of it by additional 1-form ``gauge fields'' $A$ such that the resulting action functional is a gauging of the original one. Traditionally, these gauge fields are valued in some Lie algebra $\mf{g}$, or from an alternative point of view, there exists an action Lie algebroid $M\times \mf{g}$ and the action of $\mf{g}$ on the manifold $M$ foliates it into gauge orbits (leaves of the foliation). However, as discussed in Ref. \cite{Kotov:2014iha,ChatzistavrakidisAHP}, this question can be posed and answered for a much wider class of singular foliations that do not result from a group action. One should note that general singular foliations are the rule rather than the exception. They come from $L_{\infty}$-algebroids rather than just Lie algebroids or even action Lie algebroids; see \cite{foliations} for more details. 

Given a singular foliation ${\cal F}$ on $M$, the question becomes whether an action functional ${\mc S}_0[X,A]$ exists{\footnote{With an outlook towards section \ref{sec3}, we have denoted this classical action functional with a subscript 0, which later on will indicate that it does not contain any antifields.}} such that when $A=0$ it reduces to \eqref{sigmamodel} and such that it has a gauge symmetry which on the scalar fields corresponds to deformations along the leaves of ${\cal F}$. Indeed, under relatively mild assumptions \cite{Grgur} the general form of this action functional turns out to be 
\bea \label{ugt}
{\cal S}_0[X,A]=-\int_{\S} \left(\frac 12\, g_{\m\n}(X) F^{\m}\w\ast\, F^{\n}+A^a\w\theta_a(X)+\frac 12 \g_{ab}(X)A^a\w A^b\right)-\int_{\widehat{\Sigma}}X^{\ast}H\,,  
\eea   
where $\theta_a=\theta_{a\m}(X)\dd X^{\m}$ and $\g_{ab}$ arise as  pullbacks of a collection of 1-forms and functions on $M$, respectively. 
The covariant exterior differentials, that are denoted by $F^{\m}$ in this action functional, are defined as 
\be 
F^{\m}:=\dd X^{\m}-\rho^{\mu}_{a}(X)A^{a}\,  \label{F}
\ee 
where $\rho^{\mu}_a$ are the $X$-dependent components of a host of vector fields that generate the foliation ${\cal F}$. For a consistent gauging one needs to specify the data $\rho^{\mu}_a$, $\theta_a$, and $\gamma_{ab}$ together with where the gauge fields $A$ take values in. This was considered in detail in \cite{ChatzistavrakidisAHP}. 

Here, in the context of \emph{topological} Dirac sigma models, we focus on the subclass of cases where the singular foliation comes from a (maximal rank) Dirac structure in an exact Courant algebroid on $M$ with \v{S}evera class $[H]$. These are, among others, particular Lie algebroids $E \to M$ whose rank equals the dimension of $M$. The gauge field 1-forms $A$ take values in $E$, $A=A^{a}\otimes e_a \in \Omega^{1}(\S;X^{\ast}E)$, where $e_a$ denotes a local basis in $\Gamma(E)$ as well as in $\Gamma(X^*E)$, depending on the context.  As all  Lie algebroids, Dirac structures come together with an anchor map
\be  
E \overset{\rho}\longrightarrow TM\,,
\ee 
whose image generates the foliation ${\cal F}$. Note that its components $\rho^{\m}_{a}$ do not form an invertible matrix (except in the uninteresting case where one gauges all of $M$).

Being equipped with a Lie algebroid structure, $E$ is endowed with a Lie bracket $[\cdot,\cdot]_{E}$ on its sections. In a local basis $e_a$ this gives rise to ($x$-dependent) structure functions $C_{ab}^{c}$, which also  govern the involutivity of the generating vector fields $\rho_{a}=\rho(e_{a})=\rho^{\m}_{a} \partial_\mu$, 
\be \label{closure}
[e_a,e_b]_{E}=C_{ab}^{c}e_{c} \quad \Rightarrow \quad [\r_a,\r_b]=C_{ab}^{c}\r_{c}\,.
\ee  
The Lie bracket $[\cdot,\cdot]_{E}$ satisfies the Jacobi identity 
\be \label{Jacobi}
[[e,e']_{E},e'']_{E}+\text{cyclic}=0\,, \quad \forall \: e,e',e''\in \G(E)\,.
\ee    
We note that \eqref{closure} and \eqref{Jacobi} lead to the following component identities for the structure functions
\bea
&&2\rho^{\n}_{[a}\partial_{\n}\rho^{\m}_{b]}=C_{ab}^{c}\rho^{\m}_{c}\,, \label{closurecomp} \\[4pt]
&&\rho^{\m}_{[a}\partial_{\m}C^{d}_{bc]}=C^{d}_{e[a}C^{e}_{bc]}\,, \label{Jacobicomp}
\eea 
which will be used extensively below. 

The Dirac structure $E$ enters the gauged action functional in the following way: Choosing a  basis in $E \subset TM \oplus T^*M$, $e_a =  \rho_a + \theta_a$, we obtain the anchor entering \eqref{F} as well as the 1-forms $\theta_a$, which couple linearly to the gauge fields in \eqref{ugt}. From these two quantities one determines the functions $\gamma_{ab}$ by contraction: 
$\gamma_{ab} = \iota_{\rho_a}  \theta_b$.  Note that these functions are automatically antisymmetric in their indices $a$ and $b$  due to the isotropy condition  holding true for Dirac  structures,  $\iota_{\rho_a}  \theta_b + \iota_{\rho_b}  \theta_a=0$. 

It is remarkable that there are \emph{no} condititions needed  for  $g$ and $H$ so that \eqref{ugt} is the  desired gauge theory. Still, one needs to specify  the gauge transformations to make this explicit. We make a small detour to recall the corresponding details, also because the gauge transformations are an essential ingredient for the BV theory and, in the case at hand, are not so easy  to find.

\subsection{Nonstandard gaugings and the gauge transformations}
\label{sec22}

We still need to say under which conditions on the generalized metric $V \subset TM \oplus T^*M$, determined by $g$ and $H$, a gauging along the Dirac structure $E$ is possible. For this purpose \cite{Severa:2019ddq}, one considers the subbundle $V_E := E \oplus (E^\perp \cap V)$, where $V=\{ v + \iota_v g,  v \in TM \}$ and $E^\perp$ is the orthogonal to $E$ with respect to the canonical inner product on the generalized tangent bundle $TM \oplus T^*M$. Note that one has $E \subset E^\perp$, which is possible due to the indefinite signature. The condition found in \cite{Severa:2019ddq} for that one can perform such a gauging is 
\be [\Gamma(V_E),\Gamma(V_E)] \subset \Gamma(V_E)\, . \label{SeSt}
\ee 
Now in the topological context, $E$ has maximal rank and, since $E=E^\perp$ and thus $V_E=E$, this condition is automatically satisfied by the requirements that $E$ is a Dirac structure. 

This is at first sight somewhat astonishing, the gauging along a (full rank) Dirac structure can be effectuated for \emph{every} choice of $g$ and $H$---and it was in fact this observation that led to \cite{Kotov:2014iha}. If the bundle $E$ is an involutive and isotropic subbundle of $TM \oplus T^*M$, but not of maximal rank, then the condition \eqref{SeSt} becomes a non-empty one for the metric, the 3-form, and the data determining $E$. More generally, one may even consider gaugings where one has a bundle $\widehat{E}$ over $M$ with a local basis $\widehat{e_a}$  together with a map $\widehat{\rho} \colon \widehat{E} \to TM \oplus T^*M$, $(x,\widehat{e_a})\mapsto (x,\rho_a + \theta_a)$ such that the image does not have constant rank but defines a smooth submodule of the sections; this was performed in \cite{ChatzistavrakidisAHP} and for the resulting conditions one may consult that paper (the conditions are recalled in a different form below, however). 

Here we are primarily interested in the topological Dirac sigma model, $E$ being a Dirac structure (in the original sense, i.e.\ being defined as maximally isotropic, involutive subbundles). As we argued above, there are no conditions on the metric $g$ and the closed 3-form $H$ to be satsified for it to exist. However, for the BV formulation of the theory, we need to parametrize the generators of the gauge transformations. And, to specify them, we need additional data, which will turn out to be two connections on $M$, satisfying some compatibility conditions. To address this question, we return to the general, not necessarily topological setting, since, at the initial point, it does not make a difference.

In which sense is then the action functional \eqref{ugt} a gauging of the original one \eqref{sigmamodel}? First, evidently ${\cal S}_{0}[X,A=0]=S[X]$, which is the prime requirement as explained above. In addition, with regard to the second requirement, there is a gauge symmetry  
\bea 
\d X^{\m}&=&\r^{\m}_{a}(X)\e^{a}\,,\label{gt1} \\[4pt] \label{gt2}
\d A^{a}&=& \dd\e^{a}+C^{a}_{bc}(X)A^b\e^c+\omega^{a}_{b\m}(X)\e^bF^{\m}+\phi^{a}_{b\m}(X)\e^b\ast \! F^{\m}\,,
\eea 
where $\e^{a}\in \G(X^{\ast}E)$ is the $\s$-dependent scalar gauge parameter. 

The last two terms in the gauge transformation of $A^{a}$ are needed for gauge invariance in the general case. Albeit the fact that they are vanishing onshell---we recall that $F^{\m}=0$ is the field equation for $A^{a}$---only a part of them correspond to trivial gauge symmetries \cite{HT} (which are symmetries present for every action functional). For example, for the $H$-twisted Poisson sigma model without a kinetic term, to which we will repeatedly specialize our formulas below, the trivial gauge symmetries correspond to the choice of a torsion-free connection, to which, however, one needs to add a torsion term. 

Now let us just address the question, under which conditions on all the coefficients in an action functional of the form \eqref{ugt} and  transformations of the form \eqref{gt1} and \eqref{gt2}  one obtains a gauge invariance of the functional. 
There are four conditions one finds \cite{ChatzistavrakidisAHP}. The first two do not contain the metric. First, $\gamma_{ab}$ needs to equal $\iota_{\rho_{a}}\theta_b$ and, given that $\gamma_{ab}$ is antisymmetric, this means that $\iota_{\rho_{(a}}\theta_{b)}=0$ so that the sections $\rho_a+\theta_a$ of the extended bundle $TM\oplus T^{\ast}M$ need to span an isotropic subspace.  The second condition is 
\be \label{ic3}
{\cal L}_{\rho_{a}}\theta_{b}-\iota_{\rho_{b}}\dd\theta_a-\iota_{\rho_a}\iota_{\rho_{b}}H=C_{ab}^c\theta_c\,.
\ee  
Combining this with the involutivity of the vector fields $\rho_a$, one recognizes that  $\rho_a+\theta_a$ are constrained to be also involutive. If they span a constant  subbundle $E$ of maximal rank, it means precisely that one needs to have a Dirac structure in the $H$-twisted standard Courant algebroid over $M$.

The remaining two conditions for gauge invariance read as follows: 
\bea 
{\cal L}_{\rho_a}g&=&\omega_a^{b}\vee \iota_{\rho_b}g+\phi_a^b\vee\theta_b\,, \label{ic0g} \\[4pt]
\iota_{\rho_a}H&=&\dd \theta_a-\omega_a^{b}\w\theta_b-\phi_a^{b}\w\iota_{\rho_b}g\,.\label{ic0H}
\eea 
To express these conditions in a suggestive and more geometrical form, we first define the sections 
\be \label{gop}
\mc G_{\pm}=\theta\pm \r^{\ast} \in \G(T^{\ast}M\otimes E^{\ast})\,,
\ee  
with $\r^{\ast}=\iota_{\rho_a}g\otimes e^{a}$ and $\theta=\theta_a\otimes e^{a}$.  These correspond to the projections $\pi_\pm$ in \cite{Severa:2019ddq}.

Now we note that the coefficients in front of the last two  terms in \eqref{gt2} are geometrically meaningful: $\omega^{b}_{a\m}$ are the components of a connection  $\nabla^{\omega} \colon \G(E)\to \G(T^{\ast}M\otimes E)$ on $E$ such that 
\be 
\nabla^{\omega}e_a=\omega^{b}_{a}\otimes e_b=\omega^{b}_{a\m}(x)\,\dd x^{\m}\otimes e_b\,,
\ee  and $\phi^{b}_{a\m}$ are the components of a 1-form valued endomorphism $\phi\in \G(T^{\ast}M\otimes \text{End}(E))$. This geometric nature of these quantities can be inferred from their transformation properties under field-dependent changes of basis: $\omega$ transforms inhomogeneously, as a connection, and $\phi$ homogeneously, as a tensor \cite{ChatzistavrakidisAHP,Chatzistavrakidis:2016jci}. 
Using the fact that two connections differ by an endomorphism valued 1-form, it is convenient to trade the pair $(\nabla^{\omega},\phi)$ with a pair of connections $\nabla^{\pm}:\G(E)\to \G(T^{\ast}M\otimes E)$, defined such that 
\be 
\nabla^{\pm}e_a=\Omega^{\pm}{}_a^{b}\otimes e_b\,,
\ee 
where $\Omega^{\pm}=\omega\pm \phi$. Note then that the endomorphism is obtained as the difference
\be 
\phi=\frac 12 \left(\nabla^{+}-\nabla^{-}\right)\,.
\ee 
In the following we will always use these two connections instead of $(\nabla^{\omega},\phi)$ and we will return to them and describe some of their properties in more detail in Section \ref{sec23}. 

Denoting with the same letter $\nabla^{\pm}$ the connections on tensor powers of $E$ and $TM$, induced by $\nabla^{\pm}$ on $E$ and by the Levi-Civita connection on $TM$ 
and by $\mathrm{D}^{\pm}$ the associated exterior covariant derivatives acting on the exterior algebra of $E^{\ast}$, one obtains the following generalised invariance conditions on the couplings, written in a geometric, frame-independent way:{\footnote{These should be compared to the equivalent ones originally found in \cite{ChatzistavrakidisAHP} in terms of $(\nabla^{\omega},\phi)$: 
\bea \label{ic1old}
\text{Sym}\nabla^{\omega}{\rho}^{\ast}&=&\text{Sym} \langle \phi\, \overset{\otimes}, \, \theta\rangle \,, 
\\[4pt] \label{ic2old}
\iota_{\rho}H&=& \mathrm{D}^{\omega}\theta-\langle \phi\, \overset{\w}, \, {\r}^{\ast}\rangle\,.
\eea 
}}
\bea \label{ic1}
&&\text{Sym}\left(\nabla^{+}{\cal G}_+- \nabla^{-}{\cal G}_{-}\right)=0\,, 
\\[4pt] \label{ic2}
&&  \mathrm{D}^{+}{\cal G}_++\mathrm{D}^{-}{\cal G}_{-}=2\,\iota_{\rho}H\,,
\eea 
where $\rho=\rho^{\m}_{a}\,e^{a}\otimes \partial_{\m} \in \G(E^{\ast}\otimes TM)$.   
Essentially, one may think of \eqref{ic1} as a generalised Killing equation for the metric $g$, which enters through the sections ${\cal G}_{\pm}$. 

For $E$ a Dirac structure, the background data $g$ and $H$ do not need to satisfy any conditions to render \eqref{ugt} topological. According to \eqref{SeSt} 
there always \emph{exist} connections $\nabla^+$ and $\nabla^-$ such that the transformations \eqref{gt1} and \eqref{gt2} leave the functional invariant. We will provide explicit formulas for them in the following, since these are needed for establishing the master equation in the BV formalism, where the generators of the gauge transformations enter the BV extension.

\subsection{Curvature and torsion vs. $E$-curvature and $E$-torsion}
\label{sec23}

For topological Dirac sigma models described by the action functional \eqref{ugt}, the vector bundle $E$ is identified with a Dirac structure of the $H$-twisted standard Courant algebroid on $M$. This has the advantage that the coefficients $\omega^{a}_{b\m}$ and $\phi^{a}_{b\m}$, respectively $\Omega^{\pm}{}_{b\m}^{a}$, introduced in the previous section can be calculated explicitly. The main reason is that the range of values for the Greek and Latin indices is the same, namely that the rank of $E$ is the same as the dimension of the target space $M$. Thus, although neither $\rho^{\m}_{a}$ nor $\theta_{a\mu}$ are invertible in general, there exist combinations of them that are always invertible for Dirac structures. Indeed, as proven in \cite{Chatzistavrakidis:2016jci}, the maps
\be \label{eop}
\mc{E}_{\pm}:=\theta^{\ast}\pm \rho: E\to TM\,,
\ee 
where $\theta^{\ast}=g^{-1}(\theta,\cdot)$, $g^{-1}$ being the inverse of the nondegenerate Riemannian metric $g$, are invertible. In other words, one could think, at least with some caution, of $\mc E$ with components $\mc E^{\m}_{a}$ as a generalised (inverse) Vielbein. 
Accordingly, the sections ${\cal G}_{\pm}$, thought of as maps from $E$ to $T^{\ast}M$, are also invertible since ${\mc G}_{\pm}=g\,\mc E_{\pm}$ and they have the advantage that the limit $g\to 0$ can be taken directly. For instance, this is important in comparing all results to the $H$-twisted Poisson sigma model without a metric term.  

The invertibility of the maps ${\cal E}_{\pm}$, respectively ${\cal G}_{\pm}$, has the following consequence on the equations of motion of Dirac sigma models. Variation of the action \eqref{ugt} with respect to the 1-form $A^{a}$ yields the field equations
\be 
 \left(\theta_{a\m}-(\iota_{\rho_{a}}g)_{\m}\ast\right) F^{\m}=0\,. 
\ee 
where $\gamma_{(ab)}=0$ was used. For Dirac structures, the parenthesis corresponds to an invertible operator (which remains true with the appearance of the Hodge $\ast$ in the second term, see \cite{Kotov:2004wz}) and therefore the field equation is simply 
\be 
F^{\m}=0\,.
\ee 
 We also note that the field equation for the scalars $X^{\m}$ reads 
\be \label{Xeom}
G_{\m}:=\dd(\theta_{a\m}A^{a})+\frac 12 \left(\rho^{\n}_{b}\partial_{\m}\theta_{a\n}-\theta_{a\n}\partial_{\m}\rho^{\n}_{b}+H_{\m\n\r}\rho^{\n}_{a}\rho^{\r}_{b}\right)A^{a}\w A^{b}=0\,.
\ee 
We revisit this equation from a target space covariant perspective in Section \ref{sec24}. 

With the above definitions and for an arbitrary Dirac structure $E$, a possible choice for the coefficients $\omega^{a}_{b\m}$ and $\phi^{a}_{b\m}$  can be provided in a closed form \cite{Chatzistavrakidis:2016jci}. It is more useful though to directly present expressions in terms of the coefficients $\Omega^{\pm a}_{b\m}$ of the connections $\nabla^{\pm}$ on $E$:
\be 
\Omega^{\pm}{}_{b\m}^{a}=(\mc G_{\pm}^{-1})^{a\n}\left(\,\partial_{\m}{\mc G}_{\pm}{}_{b\n}-\mathring{\Gamma}_{\n\m}^{\rho}{\cal G}_{\pm b\rho}- \frac 12 \,\r^{\s}_{b}H_{\mu\n\s}\right)\,.
	\ee 
Here	$\mathring{\Gamma}^{\rho}_{\m\n}$ are the coefficients of \emph{any} torsion-free connection on M, e.g.\ the Levi-Civita connection. For every such a choice one satisfies both equations, Eq. \eqref{ic1} and Eq. \eqref{ic2}.

	At this stage, it is advantageous to think of the connections $\nabla^{\pm}$ as being induced by some auxiliary connection $\nabla$ on $TM$. In case $E=T^{\ast}M$ this is simply the dual connection $\nabla^{\ast}$ on the cotangent bundle which is induced naturally by the one on the tangent bundle through the following equation for the corresponding covariant derivatives 
	\be 
	\langle\nabla^{\ast}_{v'}\a,v\rangle+\langle\a,\nabla_{v'}v\rangle=\dd \langle\a,v\rangle(v')\,,
	\ee 
	where $v,v'\in \G(TM)$, $\a\in \G(T^{\ast}M)$ and the angle brackets denote the natural pairing of the tangent and cotangent bundles. Recall that in a coordinate basis where the $TM$ covariant derivative has coefficients 
	\be 
	\nabla_{\m}\partial_{\n}=\Gamma^{\rho}_{\n\m}\partial_{\rho}\,,
	\ee 
	the dual covariant derivative becomes 
	\be 
	\nabla_{\m}^{\ast}\dd x^{\n}=\G^{\ast}{}^{\n}_{\rho\m}\dd x^{\rho}=-\G^{\n}_{\rho\m}\dd x^{\rho}\,.
	\ee 
	 However, in the present discussion we would like to work with the general vector bundle (Dirac structure) $E$. In that case one should think as follows. Let the covariant derivatives of the connections $\nabla^{\pm}$ be 
	\be 
	\nabla_{v}^{\pm}:\G(E)\to \G(E)\,,
	\ee         
	satisfying the usual linearity properties and Leibniz rule of covariant derivatives. We use the maps ${\cal G}_{\pm}:E\to T^{\ast}M$ given in \eqref{gop} to define covariant derivatives on $T^{\ast}M$ as follows:
	\be \label{nablatm}
	\nabla^{\ast \pm}_{v}:={\cal G}_{\pm}\circ \nabla^{\pm}_{v}\circ {\cal G}_{\pm}^{-1}: \G(T^{\ast}M)\to \G(T^{\ast}M)\,. 
	\ee 
 Now suppose that in a local coordinate basis of $T^{\ast}M$ these covariant derivatives have coefficients $\G^{\ast\pm}{}^{\m}_{\n\rho}$.
Since we already know that $\nabla^{\pm}_{\mu}e_b=\Omega^{\pm}{}^{a}_{b\m}e_a$, using the definition \eqref{nablatm}, we obtain
\be 
\G^{\ast\pm}{}^{\m}_{\n\rho}=-\mathring{\Gamma}^{\m}_{\n\rho}+\frac 12 {\cal G}_{\pm}^{-1}{}^{\m a}\rho^{\s}_{a}H_{\s\n\rho}\,.
\ee  
We can now introduce connections $\nabla^{\sharp\pm}$ with torsion on $TM$ as the dual connections to $\nabla^{\ast\pm}$. According to the above, in a local coordinate basis they have the form 
$\nabla^{\sharp\pm}_{\partial_{\m}}=\Gamma^{\pm}{}_{\m\n}^{\r}\dd x^{\n}\otimes\partial_{\rho}$
with components 
\be \label{gammas}
\G^{\pm}{}^{\r}_{\m\n}=\mathring{\G}^{\r}_{\m\n}-\frac 12 {\cal G}_{\pm}^{-1}{}^{\rho a}\rho^{\s}_{a}H_{\s\m\n}\,.
\ee 
These are indeed connections with torsion on $TM$; we denote the torsion tensors as $\Theta^{\pm}$ with components 
\be \label{thetacomp}
\Theta^{\pm}{}^{\rho}_{\m\n}:=-\G^{\pm}{}^{\rho}_{\m\n}+\G^{\pm}{}^{\rho}_{\n\m}={\cal G}_{\pm}^{-1}{}^{\rho a}\rho^{\s}_{a}H_{\s\m\n}\,.
\ee  
In more geometric terms these are two tensors $\Theta^{\pm}\in\G(TM\otimes T^{\ast}M\otimes T^{\ast}M)$ given as 
\be 
\Theta^{\pm}=\langle \rho_{{\cal G}_{\pm}}, H\rangle\,,
\ee 
where $\rho_{{\cal G}_{\pm}}={\cal G}_{\pm}^{-1}{}^{a}\otimes \rho_a \in\G(TM\otimes TM)$ and the contraction is between the second factor of this 2-vector and the first one of the 3-form $H$. 

Before proceeding further, it is useful at this stage to compare the above formulas with the ones of the $H$-twisted Poisson sigma model \emph{without} kinetic term. To achieve this, one should consider $E=T^{\ast}M$ and make the following identifications \cite{Chatzistavrakidis:2016jci}: 
\be \label{reduce}
\rho=\Pi^{\sharp} \quad \text{and}\quad  \theta=\text{id}\,,
\ee 
or, in components, $\rho^{\m}_{a}\to \Pi^{\n\m}$ and $\theta_{a\m}\to \d_{\m}^{\n}$. In addition, we would like to take a limit where the tensor $g$ goes to zero. Then from definition \eqref{gop} we  see that ${\cal G}_{+}$ and ${\cal G}_{-}$ become identical, since $\rho^* \to 0$, and moreover they are simply the identity operators: ${\cal G}_{\pm}=\text{id}$. In effect, the coefficients $\Omega^{\pm}$ also become identical and therefore they correspond to only a single connection, in other words $\phi=0$. From \eqref{gammas} we directly infer that the dual connection on $TM$ has components 
\be \label{gammaT*}
\G^{\rho}_{\m\n}=\mathring{\Gamma}^{\rho}_{\m\n}-\frac 12 \Pi^{\rho\s}H_{\s\m\n}\,,
\ee 
as expected. The coefficients $\mathring{\Gamma}^{\rho}_{\m\n}$ are of an arbitrary affine connection without torsion on $TM$. 

This result can be also directly inferred already from Eqs. \eqref{ic0g} and \eqref{ic0H} in the case $g \to 0$ together with \eqref{reduce}: The first one is identically satisfied, whereas the second one fixes the antisymmetric components, while the symmetric ones can be arbitrary. The torsion is then $\Theta^{\rho}_{\m\n}=\Pi^{\rho\s}H_{\s\m\n}$ or, in geometric terms, $\Theta=\langle\Pi,H\rangle$, which reproduces correctly the special case under consideration, see  \cite{Ikeda:2019czt}. 

We note that in the present context it is worth examining the other extremal case, which is obtained when $E=TM$. The latter is a Dirac structure only when $H=0$. Its Lie algebroid structure is given by the ordinary Lie bracket of vector fields and the choice $\rho=\text{id}$. On the other hand, $\theta=0$ and thus in agreement with \eqref{ic0H} we can consistently set $\phi=0$ as well. The two connections are once more identical then and there is only one connection left in the problem. On the other hand, the invariance condition \eqref{ic0g} introduces a further ambiguity. Specifically one finds that 
$\omega_{ij}^k=\G_{ij}^k$, where $\G^{k}_{ij}$ are the coefficients of a connection $\nabla$ on $TM$ that satisfies 
\be 
\nabla_k g_{ij}-g_{il}T_{jk}^l-g_{jl}T^{l}_{ik}=0\,,
\ee  
with $T^{i}_{jk}$ the torsion of $\nabla$. This condition is obviously satisfied for the Levi-Civita connection, but also by non-metric compatible connections with torsion. This ambiguity was also noticed before in \cite{Hancharuk:2021vxk}. Hence we are going to work with the Levi-Civita connection though. 

 Returning to the general case, aside the torsion tensors $\Theta^{\pm}$, one can define the curvature of the connections $\nabla^{\sharp\pm}$ through the usual expression
 \be 
 R^{\pm}(v,u)=[\nabla^{\sharp\pm}_{v},\nabla^{\sharp\pm}_u]-\nabla^{\sharp\pm}_{[v,u]}\,, \quad v,u\in \G(TM)\,.
 \ee 
 The analogous equation for the connections $\nabla^{\pm}$ on $E$ yield the corresponding curvature, which can also be defined as the 2-form
 \be 
 R^{\pm a}{}_b=\dd\Omega^{\pm a}{}_b+\Omega^{\pm a}{}_{c}\w\Omega^{\pm c}{}_{b}\,,
 \ee   
 as usual in the Cartan formalism. The curvature satisfies the  Bianchi identity 
\be \label{id0}
\nabla^{\pm}_{[\m}R^{\pm a}{}_{b \n\rho]}+\Theta^{\pm}{}^{\s}_{[\m\n}R^{\pm a}{}_{b \r]\s}=0\,,
\ee 
where the square brackets indicate an antisymmetrization of the enclosed indices. 

There also exist two further notions of curvature and torsion, distinct from the above, that will play a central role in the following. They are related to the concept of $E$-connections and $E$-covariant derivatives. This is a generalization of the ordinary vector bundle connection in that the vector field $v$ in the ordinary covariant derivative $\nabla_{v}$ is replaced by a section $e$ of the Lie algebroid's vector bundle $E$. In general this notion can be defined with respect to any other vector bundle $\hat E$, in which case the $E$-covariant derivative on $\hat E$ is a map 
\be 
^{E}\nabla_{e}:\G(\hat E)\to \G(\hat E)
\ee  
such that it is linear and satisfies the following Leibniz rule:
\be 
^{E}\nabla_{e}(f \hat e)=f\, ^{E}\nabla_{e}\hat e+\rho(e)f \cdot \hat e\,, \quad e\in\G(E), \hat e\in \G(\hat E)\,.
\ee 
In general, one may use this concept to define an $E$-curvature through
\bea 
\label{curvE} ^{E}R(e,e')&=&[^{E}\nabla_{e},^{E}\nabla_{e'}]- \, ^{E}\nabla_{[e,e']}\,.
\eea 
In addition, in the special case where $\hat E=E$, there also exists a notion of $E$-torsion $^{E}T\in\G(E\otimes\bigwedge^{2}E^{\ast})$, which is defined through 
\be 
\label{torE} ^{E}T(e,e')= \, ^{E}\nabla_{e}e'- \,^{E}\nabla_{e'}e-[e,e']\,.
\ee 
Finally, instead of working with the $E$-curvature $^{E}R$ it often becomes advantageous to introduce the tensor 
\be 
S=\nabla (^{E}T)+2\text{Alt} (\iota_{\rho}R)\,,
\ee 
which is called the basic curvature in \cite{Ikeda:2019czt}. In the following we drop the indicator $E$ from the notation of the $E$-torsion, hence $^{E}T\equiv T$. 

The reason we introduced the above jargon is that the ordinary connections $\nabla^{\pm}$ on $E$ give rise to two $E$-connections on $E$ via the identification 
\be 
^{E}\nabla^{\pm}_{e}e':=\nabla^{\pm}_{\rho(e)}e'\,.
\ee 
This allows us to define the corresponding $E$-torsions 
$T^{\pm}$, whose components take the form
\bea \label{Etorsion}
{T}^{\pm c}{}_{ab}=-C^{c}{}_{ab}+2\iota_{\rho_{[a}}\Omega^{\pm c}_{b]}\,,
\eea
in terms of the structure functions of the vector fields $\rho_a$ and the connection coefficients $\Omega^{\pm}$. Similarly, there are two basic curvatures $S^{\pm}$, one for each $E$-connection, which have the following component form
\be 
S^{\pm}{}^{a}{}_{bc}=\nabla^{\pm}T^{\pm}{}^{a}{}_{bc}+2\iota_{\rho_{[b}}R^{\pm}{}^{a}{}_{c]}\,,
\label{spm}
\ee 
with $S^{\pm}{}^{a}{}_{bc}=S^{\pm}{}^{a}{}_{bc\m}\dd x^{\m}$.
These $E$-torsions and basic curvatures will play a crucial role in determining the form of the BV action for Dirac sigma models in Section \ref{sec3}. Notably, they satisfy a number of useful identities, which we present in component form:{\footnote{Note that these are the generalizations for arbitrary Dirac structure $E$ of equations (D.11)-(D.14) of Ref. \cite{Ikeda:2019czt} for the Dirac structure on $T^{\ast}M$, to which they reduce under the special choice \eqref{reduce} and $g \to 0$.}}
\bea 
\label{id1}{T}^{\pm c}{}_{ab}\r_c^{\m}&=& -2\rho^{\n}_{[a}\nabla^{\pm}_{\n}\r^{\m}_{b]}+\r_{a}^{\n}\r_{b}^{\s}\Theta^{\pm}{}^{\m}_{\n\s}\,,
\\[4pt]
\label{id2}\iota_{\rho_{[a}}\iota_{\rho_{b}}R^{\pm d}{}_{c]}&=&\r^{\m}_{[a}\nabla^{\pm}_{\m}{T}^{\pm d}{}_{bc]}+{T}^{\pm d}{}_{e[a}{T}^{\pm e}{}_{bc]}\,,\\[4pt]
\label{id3} [\nabla^{\pm}_{\m},\nabla^{\pm}_{\n}]{T}^{\pm a}{}_{bc}&=&\Theta^{\pm}{}^{\r}_{\m\n}\nabla^{\pm}_{\r}{T}^{\pm a}{}_{bc}+{T}^{\pm d}{}_{bc}R^{\pm a}{}_{d\m\n}-{T}^{\pm a}{}_{dc}R^{\pm d}{}_{b\m\n}-{T}^{\pm a}{}_{bd}R^{\pm d}{}_{c\m\n}\,,\,\,\,\,\,\,\,
\\[4pt]
\label{id4} T^{\pm}{}^{c}_{ab}&=&
{\cal G}_{\pm}^{-1}{}^{c\rho}
\left(\rho^{\n}_{[a}\nabla^{\pm}_{\rho}\theta_{b]\n}-
\theta_{[b \n}\nabla^{\pm}_{\rho}\rho^{\n}_{a]}-\rho^{\m}_{[a}\theta_{b]\n}\Theta^{\pm}{}^{\n}_{\m\rho}
\right)\,,
\eea 
where in the cases when the covariant derivative acts both on tangent and bundle indices, it is the covariant derivative combining the ones on $TM$ and $E$, $\nabla^{\sharp\pm}_{v}$ and  $\nabla^{\pm}_{v}$, respectively.  
For the reader's orientation we note that the first identity is proven by covariantization of the fundamental equation \eqref{closure} or  \eqref{closurecomp}, the second identity is proven by covariantization of the Jacobi identity \eqref{Jacobi} or \eqref{Jacobicomp}, the third is a direct consequence of the definitions and it is thus proven by straightforward calculation of the left-hand side, and finally the fourth identity is proven using \eqref{ic3}, \eqref{ic1} and \eqref{ic2}.

\subsection{Target space covariant formulation}
\label{sec24} 

Previously we presented the action, the classical field equations, and the gauge transformations of the Dirac sigma model in a form such that, although spacetime covariance is manifest, the target space covariance is not. However, the model is defined not only on a local patch of the target space but globally; here we explain how target space covariance is guaranteed via the two connections $\nabla^{\pm}$ and how a basis-independent formulation can be achieved.

Let us first examine the gauge transformations of the model, given in Eqs. \eqref{gt1} and \eqref{gt2}. The first one is already accounted for by previous basis-independent definitions and reads 
\be 
\d X=\rho(\epsilon)\,.
\ee 
As for the gauge transformation of the $X^{\ast}E$-valued 1-form, one should note that $A=A^{a}\otimes e_a$ in a local basis $(e_a)$ of the vector bundle $X^{\ast}E$. Noting that any of the connections $\nabla^{\pm}$ can account for the change of frame due to a change of the base point, it holds that{\footnote{Depending on which connection one uses to perform this operation, the meaning of the symbol $\d$ of the transformation changes; one could use the notation $\d^{\pm}$ to indicate this. However, we will not denote this explicitly here, since it should be clear from the context.}} $\d e_{a}=\Omega^{\pm}{}_{a\m}^{b}\d X^{\m}e_{b}$. 
Then one finds that 
\be 
\d A=(\d A_a+\rho^{\m}_{c}\Omega^{\pm}{}^{a}_{b\m}A^{b}\e^{c})\otimes e_{a}\,.
\ee   
On the other hand, starting from \eqref{gt2} and using the definitions of $\Omega^{\pm}$ and the formulas \eqref{Etorsion} for the $E$-torsions, one finds 
\be 
\d A^{a}=\DD^{\pm}\epsilon^{a}-T^{\pm}{}^{a}_{bc}A^{b}\epsilon^{c}-\rho^{\m}_{c}\Omega^{\pm}{}^{a}_{b\m}A^{b}\epsilon^{c}+\frac 12 \left(\Omega^{+}{}^{a}_{b\m}(1+\ast)+\Omega^{-}{}^{a}_{b\m}(1-\ast)-\Omega^{\pm}{}^{a}_{b\m}\right)\epsilon^{b}F^{\m}\,.
\ee 
These are two equations, depending on the choice of connection in the covariantization. Since both connections are important in the present context, it is useful to average over these equations and obtain 
\be \label{dAtensor}
\d A=\frac{\DD^{+}+\DD^{-}}{2}\,\epsilon-\frac {T^{+}+T^{-}}{2}\,(A,\epsilon)+\biggl\langle\frac {\Omega^{+}-\Omega^{-}}{2},\ast F\biggl\rangle(\epsilon)\,.
\ee  
Recall that $(\Omega^{+}-\Omega^{-})/2$ is nothing but the tensor $\phi$. This is then the tensorial transformation rule of the $X^{\ast}E$-valued 1-form $A$. It is useful to examine what happens in the special case of $E=T^{\ast}M$ with the $H$-twisted Poisson sigma model choices \eqref{reduce} and $g \to 0$. As discussed in Section \ref{sec22}, then the two connections become identical, $\DD^{+}=\DD^{-}\equiv \DD$, which also implies that $T^{+}=T^{-}\equiv T$, and the tensor $\phi$ vanishes. In that case, the transformation of $A$ given in \eqref{dAtensor} reduces to
\be 
\d A|_{\text{HPSM}}=\DD\epsilon-T(A,\epsilon)\,,
\ee 
as found in \cite{Ikeda:2019czt}. 

Let us now move on to the field equations of the Dirac sigma model. The one for $A$ is essentially already in a target space covariant form, specifically 
\be 
F:=\dd X-\rho(A)=0\,.
\ee  
Recall that $\dd X$ is a linear map from $T_p\S$ to $T_{X(p)}M$ for every $p\in \S$ and the above equation is well defined. The field equation for the scalar fields $X^{\m}$, given in \eqref{Xeom}, may be covariantized using the connections $\nabla^{\pm}$. Using the identity \eqref{id4} and defining $\mf{a}:=\theta(A)$, we find the covariant expression{\footnote{Either of the two equivalent equations with $\pm$ can be used, or alternatively one could sum the two and obtain a more democratic expression.}} 
\be 
G_{\m}=\DD^{\pm}\mf a_{\m}-\frac 12 ({\cal G}_{\pm}{}_{c\m}T^{\pm}{}^{c}{}_{ab}\pm{\cal G}_{\pm}^{-1}{}^{\rho c}(\iota_{\rho_{a}}g)_{\rho}(\iota_{\rho_c}\iota_{\rho_b}H)_{\m})A^{a}\w A^{b}\,.
\ee 
The basis-independent form of this field equation may then be written as 
\be 
G=\DD^{\pm}\mf{a}-\frac 12 \, T^{\pm}_{{\cal G}_{\pm}}(A,A)\mp \frac 12 \, \Theta_{\ast}^{\pm}(\cdot,\rho(A),\rho(A))\,,
\ee 
where $T^{\pm}_{{\cal G}_{\pm}}={\cal G}_{\pm a}\otimes T^{\pm a}$ is a section of $T^{\ast}M\otimes \bigwedge^2 E^{\ast}$ and we have defined $\Theta_{\ast}^{\pm}$ as the contraction of the torsion tensors $\Theta^{\pm}$ with the metric, having components $\Theta^{\pm}_{\ast\, \m\n\rho}=\Theta^{\pm \s}_{\m\n}g_{\rho\s}$. 
 For the special case of the $H$-twisted Poisson sigma model without kinetic term, where all $\pm$ quantities become identical, the metric $g$ vanishes and ${\cal G}$ and $\theta$ are the identity operators, this equation reduces to the simpler expression
\be 
G|_{\text{HPSM}}=\DD A-\frac 12 \, T(A,A)\,.
\ee 

Finally, regarding the action functional \eqref{ugt}, one may consider the maps (denoted by the same name as the corresponding sections for brevity) $\rho \colon E\to TM$ and $\theta \colon E\to T^{\ast}M$ and express it as 
\be 
{\cal S}_{0}[X,A]=-\int_{\S}\left(||F||^{2}+\bigg\langle(\theta\circ X)(A),\dd X+\frac 12 (\rho\circ X)(A)\bigg\rangle\right)-\int_{\widehat{\S}}X^{\ast}H\,,
\ee 
where $||F||^{2}:=(g\circ X)(F\overset{\w},\ast F)$ and where $\circ X$ denotes composition with the map $X$.

\section{The BV action with two connections}
\label{sec3} 

\subsection{The ``Q versus QP problem''}
\label{sec31}

Our main goal in this section is to construct the BV-BRST action for (topological) Dirac sigma models given by the classical action functional \eqref{ugt}. As discussed in the Introduction, there exists a powerful systematic method for determining the BV action for a topological field theory that satisfies the classical master equation, which goes under the name of the AKSZ construction \cite{Alexandrov:1995kv}. This method resides on the geometric concept of QP manifolds. These are differential graded (super)manifolds, meaning that one has $\mathbb{Z}$-graded manifolds with the induced $\mathbb{Z}_2$-grading equipped with a cohomological vector field Q (a Q structure) and, in addition, they are endowed with a graded symplectic structure (a P structure) which is Q-invariant, or in other words the Q and P structures are compatible. Once this QP structure is given, one just applies the general AKSZ algorithm to find a solution to the classical master equation---for the classical action one constructs at the same time.

There are two potential reasons that the above method would not be applicable. The first is that although the target space may carry both a Q and a P structure, these structures might not be compatible. Such a situation typically arises in presence of Wess-Zumino terms, which present obstructions to QP-ness, as e.g. in the $H$-twisted Poisson sigma model \cite{Ikeda:2019czt} and higher dimensional generalizations thereof \cite{Ikeda:2013wh,Chatzistavrakidis:2021nom,Ikeda:2021rir}. The second and more radical reason is that the Q manifold at hand might not even admit a natural symplectic structure, for example when it is not a (graded) cotangent bundle. Two-dimensional Dirac sigma models constitute a class of topological field theories where both issues described above appear. Indeed, in general there is no natural P structure in this context, apart from very special cases such as when $E=T^{\ast}M$. Even in such cases though, QP-ness is obstructed by the 3-form $H$. 

Let us now offer a more technical account for the above statements. First of all, in the framework we have employed, there always exists a natural Q structure. This is true due to Vaintrob's formulation of Lie algebroids as homological vector fields on degree-shifted vector bundles \cite{Vaintrob} and the fact that Dirac structures are Lie algebroids. A Lie algebroid $(E,[\cdot,\cdot]_{E},\rho:E\to TM)$ over $M$ with its Lie algebra structure on the sections of the vector bundle $E$ and the anchor map $\rho$ may alternatively be described by the parity-reversed vector bundle $E[1]$, equipped with local coordinates $x^{\m}$ and $\xi^{a}$ of degrees zero and one, respectively. Consider the odd vector field (in fact, the most general one of degree one is of this form always)
\be 
Q_{E}=\rho^{\m}_{a}(x)\xi^{a}\partial_{x^{\m}}-\frac 12 C^{a}_{bc}(x)\xi^{b}\xi^{c}\partial_{\xi^{a}}\,,
\ee 
where $\partial_{x^{\m}}:=\partial/\partial x^{\m}$ and $\partial_{\xi^{a}}:=\partial/\partial{\xi^{a}}$.   Then $E$ is a Lie algebroid if and only if $(E[1],Q)$ is a Q manifold, namely $Q_E^{2}=0$.  More precisely, this works under the identifications $\rho({e}_a)=\rho_{a}^{i}(x)\partial_{x^{i}}$  and $[{e}_a,{e}_b]_{E}=C_{ab}^{c}(x){e}_c$ in a local basis ${e}_a$ of $\Gamma(E)$.

 Having established the existence of a Q structure, let us turn our attention to the possibility of having a compatible P structure too. We have previously defined the target space 1-form $\mf a$ as $\theta(A)$. 
 Inspecting the corresponding term in the action functional, we are prompted to define the 2-form{\footnote{One should be cautious of the fact that we have not distinguished between the pull-back field ${\mf a}$ and the degree-1 coordinate of the target space that it originates from; clearly the latter appears in the formula for $\omega$, that is ${\mf a}_{\m}=\theta_{a\m}(x)\xi^{a}$.}} 
 \be 
 {\omega}:= \mathtt{d}x^{\m}\w \mathtt{d}\mf{a}_{\m}\,,
 \ee  
 where $\mathtt{d}$ is the differential on $M$, not to be confused with the differential $\dd$ on $\S$.
 This would be the candidate graded symplectic form on the target space. First of all, it is certainly closed, $\mathtt{d}\omega=0$. However, as long as $\theta$ is not an invertible operator, it is obvious that $\omega$ is not non-degenerate. Thus, $\omega$ is presymplectic rather than symplectic.{\footnote{A discussion on presymplectic AKSZ constructions appears in Ref. \cite{Grigoriev:2020xec}.}} It becomes a genuine symplectic structure when $\theta$ is invertible, with primary example the $H$-twisted Poisson sigma model where $\theta$ is the identity operator. However, as we already mentioned this is not sufficient; $\omega$ should also be $Q_{E}$-invariant. One could test this compatibility condition already at the presymplectic level. Using the conditions of Section \ref{sec21}, we find
 \be 
 {\cal L}_{Q_{E}}\omega=\mathtt{d}\left(\frac 12 \iota_{\rho_{b}}\iota_{\rho_{a}}H\,\xi^{a}\xi^{b}\right)\,.
 \ee  
 It is directly observed that as long as the right-hand side does not vanish, and indeed it does not in all cases in the present paper, the 2-form $\omega$ is not $Q_E$-invariant, regardless if it is symplectic or presymplectic. Then the target space does not have the structure of a QP manifold and the AKSZ construction cannot be applied. 
 
Finally, the presence of the kinetic term, the first term on the right-hand side of \eqref{ugt}, is something not coming naturally from the AKSZ construction. This may change, however, if one considers boundary terms for particular boundary conditions in an AKSZ theory in one dimension higher, in which case at least some kinetic terms can be produced \cite{Severa:2016prq,Pulmann:2019vrw}; we will not pursue this path in the present paper, while it may be interesting to see if it would also lead to a valid BV formulation of the Dirac sigma model.

\subsection{BRST operator and field-antifield content}
\label{sec32}

To achieve the goal of finding the BV action for Dirac sigma models we follow the usual steps of the field-antifield formalism \cite{HT,Gomis:1994he}. Our discussion closely parallels and at the same time generalises the one of \cite{Ikeda:2019czt} for the $H$-twisted Poisson sigma model, which is a special case of the class of theories we study.  First, we enlarge the configuration space of fields by ghosts and antifields. The gauge transformation structure is promoted to BRST transformations on the fields, captured by the BRST operator. The BRST symmetry has the advantage that it remains intact even after gauge-fixing, unlike the ordinary gauge symmetry. Then we define an odd symplectic structure on the space of fields and antifields, the antibracket, and extend the classical action ${\cal S}_0$ order by order in the antifields. Finally, we show that the action we propose satisfies the classical master equation. 

The first step in implementing the field-antifield formalism is to enlarge the space of fields $\mathcal{P}=\left\{ X\colon \Sigma\rightarrow M, A\in\Omega^1\left(\Sigma, X^\ast E\right)\right\}$ of the original theory by ghosts that correspond to the gauge parameters of the ordinary gauge symmetry. We denote the new space of fields as $\mathcal{P}_{\text{BRST}}$, the ghost fields by $c^a$, and assign them ghost degree 1: $\text{gh}(c^{a})=1$. Next we define the BRST operator $s$, which is of ghost degree 1, by its action on the fields:
\begin{eqnarray}
sX^{\m} &=&\rho^{\m}_a c^a\,,\\[4pt]
sA^a &=&\dd c^a+\tensor{C}{^a_b_c}A^bc^c+\tensor{\omega}{^a_{b\m}}c^bF^{\m}+\tensor{\phi}{^a_{b{\m}}}c^b\ast F^{\m}\,,\\[4pt]
sc^a &=&-\frac{1}{2}\tensor{C}{^a_b_c}c^bc^c\,.
\end{eqnarray} 
The basic property of the BRST operator is that it should be nilpotent, at least on-shell. Indeed it is straightforward to confirm that $s^{2}X^{\m}=0$ due to the involutivity of the vector fields $\rho_a$ and that $s^{2}c^{a}=0$ due to the Jacobi identity of the Lie bracket. Finally, calculating the action of $s^2$ on the 1-form gauge fields $A^{a}$, one finds that it is not identically zero but rather proportional to terms that vanish on the classical equations of motion of the model. Specifically, the result is 
\be 
s^2A^a = \frac{1}{2}\tensor{S}{^a_b_c_\m}c^bc^cF^{\m}+\frac{1}{2}\tensor{\widetilde{S}}{^a_b_c_\m}c^bc^c\ast F^{\m}\,,
\ee 
with the two basic curvature tensors $S$ and $\widetilde{S}$  given by
\bea \label{S}
\tensor{S}{^a_b_c} &=& \frac{1}{2}\left(\tensor{S}{^{+a}_{bc}}+\tensor{S}{^{-a}_{bc}}\right)\,,\\[4pt]
\label{Stilde}
\tensor{\widetilde{S}}{^a_b_c}&=& \frac{1}{2}\left(\tensor{S}{^{+a}_{bc}}-\tensor{S}{^{-a}_{bc}}\right)\,,
\eea 
where $S^\pm$ has been defined in \eqref{spm}. The fact that $s^2$ does not vanish on all fields reflects the openness of the gauge algebra, namely that it closes only on-shell. For this reason the BRST formalism is not sufficient to construct the extended action of the classical model and one should reside to the more general BV formalism. 

The next step in implementing the BV strategy is a further extension of the field content by the introduction of antifields for each field and ghost of the theory. The new space of all the fields and antifields is simply $\mathcal{P}_{\text{BV}}=T^\ast[-1]\mathcal{P}_{\text{BRST}}$. In the present case we have three antifields $X^{+}_{\m}, A^{+}_{a}$ and $c^{+}_a$. They have the complementary form degree with respect to 2, while their ghost degree has an extra shift of $-1$, or in other words, $\text{fdeg}(\Phi^+)=2-\text{fdeg}(\Phi)$ and $\text{gh}(\Phi^+)=-\text{gh}(\Phi)-1$. We collect all fields and antifields with their ghost and form degree in Table \ref{table1}. The bi-grading of fields with respect to their ghost and form degrees $\text{gh}(\Phi)$ and $\text{fdeg}(\Phi)$, respectively, dictates their graded commutativity property, which is given by the relation 
\be 
F\w G = (-1)^{\text{fdeg}(F)\text{fdeg}(G)+\text{gh}(F)\text{gh}(G)} G\w F\,,
\ee  
for any two functionals $F$ and $G$ of the fields and antifields.

	\begin{table}
\begin{center}	\begin{tabular}{| c | c | c | c | c | c | c |}
		\hline \multirow{3}{5em}{(Anti)Field} &&&&&& \\  & $X^{\m}$ & $A^a$ & $c^a$ & $X^{+}_{\m}$ & $A^{+}_a$ & $c^{+}_a$ \\ &&&&&& \\\hhline{|=|=|=|=|=|=|=|}
		\multirow{3}{4em}{Ghost degree} &&&&&& \\  & 0 & 0 & 1 & -1 & -1 & -2 \\ &&&&&& \\\hline 
		\multirow{3}{4em}{Form degree} &&&&&& \\  & 0 & 1 & 0 & 2 & 1 & 2 \\ &&&&&&
		\\\hline 
	\end{tabular}\end{center}\caption{The classical basis with ghost and form degrees for Dirac sigma models. }\label{table1}\end{table}

The stage is now set to move on to the last part of the classical BV contruction and determine the BV action for the Dirac sigma model. We introduce a symplectic form on the space of fields and antifields:
\begin{equation}
\omega_{\text{BV}}=\int_\Sigma \left( \delta X^\mu\wedge \delta X^+_\mu +\delta A^a\wedge\delta A^+_a+\delta c^a\wedge \delta c^+_a\right)\,,
\label{symp}
\end{equation}
which induces the antibracket:
\be 
(F,G)_{\text{BV}}= \int \dd^2\sigma\, \dd^2\sigma' \sum_\Phi\left(\frac{\delta_L F}{\delta\Phi(\sigma)}\frac{\delta_R G}{\delta\Phi^\ast(\sigma')}-\frac{\delta_L F}{\delta\Phi^\ast(\sigma)}\frac{\delta_R G}{\delta\Phi(\sigma')}\right)\delta(\sigma-\sigma')\,,
\ee 
in terms of the left and right functional derivatives. Here $\Phi^\ast$ denotes the Hodge dual of the antifield $\Phi^+$. Note that the variation of any action depending on the elements of the classical basis is 
\be 
\delta S =\int \sum_\Phi \delta\Phi\frac{\delta_L S}{\delta\Phi}=\int\sum_\Phi \frac{\delta_R S}{\delta\Phi} \delta\Phi\,.
\ee 
For the field content of the Dirac sigma models with the degrees appearing in Table \ref{table1}, the complete list of nonvanishing brackets for the field components is 
\begin{align}
\left(X^{\mu}(\s),X^{+}_{01\n}(\s')\right)_{\text{BV}}&= \d^{\m}_{\n}\d^{2}(\s-\s')\,,\label{bvb1}\\[4pt]
\left(A_{0}^{a}(\s),A^{+}_{1 b}(\s')\right)_{\text{BV}}&= \d^a_b\d^{2}(\s-\s')\,,\label{bvb2}\\[4pt]
\left(A_{1}^{a}(\s),A^{+}_{0 b}(\s')\right)_{\text{BV}}&= -\d^a_b\d^{2}(\s-\s')\,,\label{bvb3}\\[4pt]
\left(c^{a}(\s),c^{+}_{01 b}(\s')\right)_{\text{BV}}&= \d^a_b\d^{2}(\s-\s')\,.\label{bvb4}
\end{align}
These explicit expressions will be particularly useful in calculations.

\subsection{BV action and the classical master equation}
\label{sec33}

The BV action $\mathcal{S}_{\text{BV}}$ needs to satisfy the classical master equation:
\begin{equation}
(\mathcal{S}_{\text{BV}},\mathcal{S}_{\text{BV}})_{\text{BV}}=0\,.
\label{cme}
\end{equation}
In general, the BV action can be expanded in the number of the antifields:
\be 
{\cal S}_{\text{BV}}={\cal S}_0+{\cal S}_1+{\cal S}_2+\ldots\,,
\ee 
where the subscripts on the right-hand side denote the number of antifields in each sector and ${\cal S}_0$ is the classical action \eqref{ugt} that contains no antifields. Moreover, the sector with one antifield is essentially fixed by the BV formalism since it should reflect the gauge invariance of the classical action. 
In particular, it is given as 
\be 
{\cal S}_{1}=\int_\Sigma\left(X_{\m}^{+}sX^{\m}-A^{+}_a\w sA^{a}-c_a^{+}sc^{a}\right)\,.
\ee 
On the other hand, the sector with two antifields is not fixed a priori and it should be such that the classical master equation is satisfied. Therefore, anticipating the final result, we make the following Ansatz for this sector of the BV action,
\be 
{\cal S}_{2}=\int_\Sigma \frac 14\left( Y^{ab}{}_{cd}(X)\,A^{+}_a\w A^{+}_b+Z^{ab}{}_{cd}(X)\, A^{+}_{a}\w\ast A^{+}_{b} \right)c^cc^d\,,
\ee   
with $Y$ and $Z$ being $X$-dependent coefficients to be determined through the classical master equation. At this stage they are arbitrary, but share the property that they are antisymmetric in their  lower two indices, whereas $Y$ is symmetric and $Z$ antisymmetric in its upper two indices, respectively. 

The next step is to impose the classical master equation \eqref{cme}. This comprises several terms with different structure that should be separately set to zero. The simplest of them is $\left({\cal S}_{0},{\cal S}_{0}\right)_{\text{BV}}$, which vanishes identically because the classical action does not contain any antifields and there is no nonvanishing component of the BV bracket that does not include antifields. Moreover, it is evident that $\left({\cal S}_{2},{\cal S}_{2}\right)_{\text{BV}}$ vanishes, even without knowing the form of $Y$ and $Z$. This is due to the fact that it contains only $A^{+}$ antifields but no $A$ fields and the only nonvanishing BV brackets of $A^{+}$ include necessarily $A$. Furthermore, a straightforward calculation establishes that $\left({\cal S}_{0},{\cal S}_{1}\right)_{\text{BV}}$ also vanishes provided the conditions \eqref{ic1}, \eqref{ic2}, \eqref{ic3} and $\gamma_{ab}=\iota_{\r_{a}}\theta_b$ hold. This is nothing but the invariance of the classical action ${\mc S}_0$ under the gauge transformations \eqref{gt1} and \eqref{gt2} implemented in the classical master equation, as expected. The next condition to be satisfied is: 
\bea 
&&\left({\cal S}_{1},{\cal S}_{1}\right)_{\text{BV}}+2\left({\cal S}_{0},{\cal S}_{2}\right)_{\text{BV}}\overset{!}=0\,.\label{cme11}
\eea 
Calculating each of the two terms results in  
\bea 
\left({\cal S}_{1},{\cal S}_{1}\right)_{\text{BV}}&=& \int  \left(\widetilde{S}^{a}{}_{cd\m}F^\mu\wedge\ast A^+_a-S^{a}{}_{cd\m}F^\mu\wedge A^+_a\right)c^cc^d\,, \\[4pt]
\left({\cal S}_{0},{\cal S}_{2}\right)_{\text{BV}}&=&\int  \frac 12 \left[\left(Y^{(ab)}{}_{cd}\,(\iota_{\rho_b}g)_\mu-Z^{[ab]}{}_{cd}\,\theta_{b\m}\right)F^\mu\wedge\ast A^+_a\right.+ \nn\\[4pt]  && \qquad\qquad \left. +\left(Y^{(ab)}{}_{cd}\,\theta_{b\m}-Z^{[ab]}{}_{cd}\,(\iota_{\rho_b}g)_\mu\right)F^\mu\wedge A^+_a\right]c^cc^d\,,
\eea 
where $S$ and $\widetilde{S}$ are given by \eqref{S} and \eqref{Stilde}, respectively. This means that imposing \eqref{cme11} fixes the two so far unknown quantities $Y$ and $Z$ in terms of the known quantities $S$ and $\widetilde{S}$. Specifically, one directly finds the inverse relations, namely those for $S(Y,Z)$ and $\widetilde{S}(Y,Z)$: 
\bea 
S^{a}{}_{cd}&=&Y^{(ab)}{}_{cd}\theta_b-Z^{[ab]}{}_{cd}\iota_{\r_{b}}g\,,
\\[4pt]
\widetilde{S}^{a}{}_{cd}&=&-Y^{(ab)}{}_{cd}\iota_{\r_{b}}g+Z^{[ab]}{}_{cd}\theta_b\,.
\eea
This prompts us to define the sum and difference of the  quantities $Y$ and $Z$, 
\be 
\tensor{Y}{^{\pm ab}_{cd}}= (Y\pm Z)^{ab}{}_{cd}\,,
\ee   
which allows us to write:
\begin{equation}
\tensor{S}{^{\pm a}_{cd}}=\tensor{Y}{^{\pm ab}_{cd}}\left(\mathcal{G}_{\mp}\right)_b\,.
\end{equation}
Due to the invertibility of the operators $\mathcal{G}_\pm$, we can invert this to express $Y^\pm$:
\be \label{Ypm}
Y_{\pm}^{ab}{}_{cd}=\langle( {\cal G}_{\mp}^{-1})^{b},S_{\pm}^{a}{}_{cd}\rangle\,.
\ee 
Then ${\cal S}_{2}$ is fully specified and the condition \eqref{cme11} holds. 

In order to find the higher order terms in the BV action, it is necessary to determine the antibracket of $\mathcal{S}_1$ and $\mathcal{S}_2$. A straightforward calculation leads to:
\bea
\left({\cal S}_{1},{\cal S}_{2}\right)_{\text{BV}}&=& \int_\Sigma  \,\frac 14 \left( I^{ab}{}_{cde}A^+_{a}\wedge A^+_{b}+J^{ab}{}_{cde}A^+_{a}\wedge\ast A^+_{b}\right)c^cc^dc^{e}\,,
\eea 
where we defined the abbreviations
\bea 
I^{ab}{}_{cde}&:=&\rho^{\m}_{[e}\partial_{\m}Y^{ab}{}_{cd]}-2(C^{(a}{}_{p[e}-\rho^{\m}_{p}\omega^{(a}_{[e\m})Y^{b)p}{}_{cd]}-2\rho^{\m}_{p}\phi^{(a}_{[e \m}Z^{b)p}{}_{cd]}-Y^{ab}{}_{p[e}C^{p}{}_{cd]},
\\[4pt]
J^{ab}{}_{cde}&:=&\rho^{\m}_{[e}\partial_{\m}Z^{ab}{}_{cd]}+2(C^{[a}{}_{p[e}-\rho^{\m}_{p}\omega^{[a}_{[e\m})Z^{b]p}{}_{cd]}+2\rho^{\m}_{p}\phi^{[a}_{[e \m}Y^{b]p}{}_{cd]}-Z^{ab}{}_{p[e}C^{p}{}_{cd]}\,.
\eea 
One now recognizes that these expressions both correspond to Bianchi identities. Indeed, since $Y$ and $Z$ are fully determined, one can first decouple these two equations by adding and subtracting them, thus expressing them in terms of $Y_{\pm}$. The latter quantities can be substituted for via \eqref{Ypm}, leading to Bianchi identities for the curvature combinations $S_{\pm}$. They are 
\bea 
&&\rho^{\m}_{[e}\nabla^{\pm}_{\m}\langle({\mc G}_{\mp}^{-1})^{b},\tensor{S}{^{\pm a}_{cd]}}\rangle-T^{\pm f}{}_{[cd}\langle ({\mc G}_{\mp}^{-1})^{b},\tensor{S}{^{\pm a}_{e]f}}\rangle+ T^{\pm a}{}_{f[e}\langle ({\mc G}_{\mp}^{-1})^{b},\tensor{S}{^{\pm f}_{cd]}}\rangle + \nn\\[4pt] && \hspace{225pt} + T^{\mp b}{}_{f[e}\langle ({\mc G}_{\mp}^{-1})^{f},\tensor{S}{^{\pm a}_{cd]}}\rangle=0\,,
\eea 
which can be proven by direct calculation using the identities \eqref{id0} and \eqref{id1}-\eqref{id4}. This shows that the antibracket of $\mathcal{S}_1$ and $ \mathcal{S}_2$ vanishes, which implies  that there are no higher order terms in the BV action needed. Therefore the BV action for the Dirac sigma model is fully determined and equal to: 
\bea 
{\cal S}_{\text{BV}}&=&-\int_{\S} \left(\frac 12 g_{\m\n}(X)  F^{\m}\w\ast F^{\n}+A^a\w\theta_a(X)+\frac 12 \g_{ab}(X)A^a\w A^b\right)-\int_{\widehat{\Sigma}}X^{\ast}H \nn\\[4pt] 
&& +\int_{\S}\left(\r_{a}^{\m}(X)X_{\m}^{+}c^{a}+\frac 12C^{a}_{bc}(X) c_a^{+}c^{b}c^{c}\right) \nn\\[4pt] 
&& -\int_{\S} A^{+}_a\w \left(\dd c^{a}+C^{a}{}_{bc}(X)A^{b}c^{c}+\omega^{a}{}_{b\m}(X)c^{b} F^{\m}+\phi^{a}{}_{b\m}(X)c^{b}\ast F^{\m}\right) \nn\\[4pt] 
&&+ \int_{\S} \frac 18 \left(\langle( {\cal G}_{-}^{-1})^{b},S^{+a}{}_{cd}\rangle(X)+\langle( {\cal G}_{+}^{-1})^{b},S^{-a}{}_{cd}\rangle(X)\right)\,A^{+}_a\w A^{+}_b c^cc^d
\nn\\[4pt]
&&+\int_{\S}\frac 18 \left(\langle( {\cal G}_{-}^{-1})^{b},S^{+a}{}_{cd}\rangle(X)-\langle( {\cal G}_{+}^{-1})^{b},S^{-a}{}_{cd}\rangle(X)\right)\, A^{+}_{a}\w\ast A^{+}_{b} c^cc^d\,.
\label{bv}
\eea 

\subsection{Manifestly target space covariant form of the BV action}
\label{sec34}

It is expected that the BV action \eqref{bv} is covariant with respect to a change of coordinates on the target space $M$. However, this covariance is not manifest due to the terms involving the antifield $X^+$. Here we aim at making this covariance manifest in the BV action.

First notice that the gauge field $A$, the ghost $c$, and their antifields $A^+$ and $c^+$ are covariant objects. Thus, if we change coordinates on $M$ and let $M^\mu_\nu(x)$ be the corresponding Jacobian matrix, with $M^a_b(x)$ the induced Jacobian matrix on $E$, the components of the field $A$ transform as
\begin{equation}
\widetilde{A}^a=M^a_b A^b\,,
\end{equation}
and similarly for the other fields.

The symplectic form \eqref{symp} needs to be invariant under the change of coordinates, or in other words, it needs to behave as a scalar. Knowing the transformations of the fields $X$, $A$, and $c$, as well as of the antifields $A^+$ and $c^+$, this leads to the transformation of the antifield $X^+$,
\begin{equation}
\widetilde{X}^+_\mu=(M^{-1})^\nu_\mu X^+_\nu-(M^{-1})^c_b\, \partial_\mu M^a_c (A^+_a\wedge A^b+c^+_a c^b)\,.
\end{equation}
This transformation property can now be used to check the covariance of the BV action.
In order to make the covariance manifest, we  covariantize the antifield $X^+$:
\begin{equation}
{X}^{+\nabla}_\mu=X^+_\mu-{\omega}^{a}_{b\mu}\left(A^+_a\wedge A^b+c^+_a c^b\right)\,,
\end{equation}
which now transforms tensorially. In terms of this field, the BV action takes the form:
\begin{eqnarray}
\mathcal{S}_{BV} &=& -\int_{\S}\left(||F||^{2}+\bigg\langle\theta(A),\dd X+\frac 12 \rho(A)\bigg\rangle\right)-\int_{\widehat{\S}}X^{\ast}H\nn\\[4pt] \label{bv2}
&& -\int_\Sigma \left(\left\langle {X}^{+\nabla},\rho(c)\right\rangle+\frac{1}{4}(T^+ +T^-) (c^{+},c,c)+\langle\phi,\ast F\rangle(A^{+},c)\right)\nn\\[4pt]
&& -\frac{1}{2}\int_\Sigma \left(\left((D^+ +D^-)c\right)(A)-(T^{+}+T^{-}) (A^{+},A,c)\right)\nn\\[4pt]
&& +\frac{1}{8}\int_\Sigma \left(\left\langle S^+(A^+,c,c),\mathcal{G}^{-1}_-(A)\right\rangle+\left\langle S^-(A^+,c,c),\mathcal{G}^{-1}_+(A)\right\rangle\right)\nn\\[4pt]
&& +\frac{1}{8}\int_\Sigma\left(\left\langle S^+(A^+,c,c),\mathcal{G}^{-1}_-(\ast A)\right\rangle-\left\langle S^-(A^+,c,c),\mathcal{G}^{-1}_+(\ast A)\right\rangle\right)\,,
\end{eqnarray}
where pull-backs via the map $X$ are understood. This is then the final covariant form of the BV action for topological Dirac sigma models. 

\section{Conclusions and outlook} 
\label{sec4}

The geometrical underpinnings of the classical master equation and its solution, the classical BV action, for topological sigma models were set in \cite{Alexandrov:1995kv} and go by the name of the AKSZ construction. The central concept in this powerful method is a QP-structure on the target space, comprising two compatible substructures, namely a cohomological vector field Q on a graded supermanifold, the Q-structure and a graded symplectic form, the P-structure. 

The AKSZ construction has its limitations. In particular, it may be the case that the target space of some topological field theory does not have the structure of a genuine QP-manifold. This can happen in a number of different ways, specifically 
\begin{enumerate}
	\item when the target space has both a Q- and a P-structure, but they are not compatible,
	\item when the would-be P-structure is degenerate and therefore presymplectic \cite{Grigoriev:2020xec}, 
	\item when the target space does not admit a P-structure at all. 
\end{enumerate} 
The prototype example of the first item in the above list is the $H$-twisted Poisson sigma model \cite{Klimcik:2001vg}, whose BV action was found in \cite{Ikeda:2019czt}. It is worth mentioning that this is certainly not an isolated example. A semi-infinite class of further examples are topological field theories with twisted R-Poisson structure in any dimension $\ge 2$ \cite{Chatzistavrakidis:2021nom}, whose BV action is also possible to identify at least in three dimensions \cite{Chatzistavrakidis:2022hlu}, although in a technically demanding way. This includes for example 4-form-twisted (pre-)Courant sigma models. A general lesson is that 
\bi 
\item in presence of Wess-Zumino terms, QP structures on the target space of topological sigma models are obstructed and the solution of the classical master equation is found by traditional methods.
\ei  
This statement is not meant to imply that the resulting solutions do not reflect the geometry of the target space. Even though a systematic construction \`a la AKSZ is still lacking in such more general settings, a lot of geometrical data are well understood in the context of higher geometry and accompanying higher notions of connections, torsion and curvature. This is also a reason that one could be confident that eventually a general and systematic method to handle Wess-Zumino terms in this spirit should exist, especially now that a variety of examples have been worked out completely. 

On the other hand, topological Dirac sigma models generically fall under item 3 of the above list. Apart from the special case of the $H$-twisted Poisson sigma model, a generic model interpolating between this special case and the completely gauged WZNW model does not have a P-structure on the target space to start with, let alone a QP one. Nevertheless, as we have shown in the present work, the solution to the classical master equation of any topological Dirac sigma model can be determined in the traditional way. We showed in particular that this BV action is given by \eqref{bv}, or in more geometric terms by \eqref{bv2}. 
Thus the second general lesson drawn in this paper is that 
\bi 
\item although a genuine QP-structure does not exist on the target space for topological Dirac sigma models, the solution to the classical master equation is fully determined. 
\ei   
As before, there is an interplay between the physical problem and higher geometry, involving two suitable Lie algebroid $E$-connections, since Dirac structures are Lie algebroids. The corresponding $E$-differential geometry was described in detail and we clarified the tensors involved in the problem ($E$-torsion and $E$-curvature), describing them also in intrinsic geometric terms. It would be interesting to explore more, possibly higher-dimensional examples and their associated geometry in this direction. In a similar spirit to the transition from 2D Poisson to R-Poisson structures in any dimension \cite{Chatzistavrakidis:2021nom}, one could generalize Dirac sigma models in higher than 2 dimensions. Some considerations in this direction are found in \cite{Ikeda:2021rir}.  

What, however, still remains open after the present work is to find the BV formulation of the non-topological Dirac sigma models, already in two dimensions. 

As a final remark, we note that it would be interesting to investigate the relation of this work to the problem of quantization. First, the solution to the quantum master equation should be constructed. Recalling on the one hand that for the ordinary Poisson sigma model the classical and the quantum BV actions do not differ (because the BV Laplacian does not contribute) \cite{Cattaneo:1999fm}  and on the other hand that a 2-point correlator on the disk corresponds to the solution of the problem of deformation quantization on a Poisson manifold \cite{Kontsevich:1997vb}, a.k.a. the star product, it should be explored how these statements extend to Dirac sigma models.

\paragraph{Acknowledgements.}  We are grateful to Zoltan Kokenyesi for discussions and participation in the early stages of the project and to Noriaki Ikeda for related discussions. This work was supported by the Croatian Science Foundation Project ``New Geometries for Gravity and Spacetime'' (IP-2018-01-7615). We have also benefited from the French-Croatian bilateral project ``Physics of non-geometric fluxes'' of the Cogito programme 2021-2022. A.Ch. and L.J. would like to thank Mariana Gra\~{n}a and Ruben Minasian for the hospitality during a visit at the IPhT-CEA Saclay where a part of this work was done.   

\appendix 
\section{Conventions}
\label{appa}

For the 2D sigma models we consider, the metric on the spacetime $\S$ is Lorentzian and fixed, specifically the 2D Minkowski metric $\eta_{\a\b}$ with signature $(-1,+1)$. Local coordinates on $\S$ are denoted as $\s^{\a}=(\s^{0},\s^{1})$. In the local basis, the Hodge operator acts as 
\be 
\ast \dd\s^{\a}=\eta^{\a\b}\epsilon_{\b\g}\dd\s^{\g}\,,
\ee   
with $\eta^{\a\b}$ the components of the inverse 2D metric and $\epsilon_{\a\b}$ the antisymmetric symbol in two dimensions with components 
\be 
\epsilon_{01}=1=-\epsilon_{10} \quad \text{and} \quad \epsilon^{01}=-1=-\epsilon^{10}\,.
\ee 
The volume form is given by 
\be 
\ast1= \dd\s^{0}\w\dd\s^{1}=\frac 12 \epsilon_{\a\b}\dd\s^{\a}\w\dd\s^{\b}\,,
\ee 
and thus 
\be 
\dd\s^{\a}\w\dd\s^{\b}=-\epsilon^{\a\b}\ast 1\,.
\ee 
In the 2D local basis, the action functional \eqref{sigmamodel} is given as 
\be 
S[X]=-\frac 12 \int \dd^2\s\left(\eta^{\a\b}g_{\m\n}\partial_{\a}X^{\m}\partial_{\b}X^{\n}+\epsilon^{\a\b}B_{\m\n}\partial_{\a}X^{\m}\partial_{\b}X^{\n}\right)-\int_{\widehat{\S}}X^{\ast}H\,.
\ee  
Note that we have set $2\pi\a'=1$ for the string slope parameter.

\end{document}